\documentclass[doublecol]{epl2}
\usepackage{graphicx}
\usepackage{amsmath}
\usepackage{bm}
\usepackage{amsfonts}
\usepackage{amssymb}

\title{First-order phase transitions in outbreaks of co-infectious diseases and the extended general epidemic process}
\shorttitle{First order phase-transitions in outbreaks of co-infectious
diseases}

\author{Hans-Karl Janssen\inst{1} \and Olaf Stenull\inst{2}}
\shortauthor{H.K. Janssen and O. Stenull}

\institute{
  \inst{1} Institut f\"{u}r Theoretische Physik III, Heinrich-Heine-Universit\"{a}t,
40225 D\"{u}sseldorf, Germany\\
  \inst{2}Department of Physics and Astronomy, University of Pennsylvania, Philadelphia,
Pennsylvania 19104, USA }


\abstract{In co-infections, positive feedback between multiple diseases can
accelerate outbreaks. In a recent letter Chen, Ghanbarnejad, Cai, and
Grassberger (CGCG) introduced a spatially homogeneous mean-field model system
for such co-infections, and studied this system numerically with focus on the
possible existence of discontinuous phase transitions. We show that their model
coincides in mean-field theory with the homogenous limit of the extended
general epidemic process (EGEP). Studying the latter analytically, we argue
that the discontinuous transition observed by CGCG is basically a spinodal
phase transition and not a first-order transition with phase-coexistence. We
derive the conditions for this spinodal transition along with predictions for
important quantities such as the magnitude of the discontinuity. We also shed
light on a true first-order transition with phase-coexistence by discussing the
EGEP with spatial inhomogeneities.}
\begin{document}

\maketitle

\section{Introduction}
The recent outbreak of ebola in Africa is the latest reminder of the
devastation that epidemic infectious diseases have caused throughout the
history of mankind. Theoretical studies of the dynamics of epidemic processes
may lead to clues for developing effective countermeasures. Such studies have
focused in the past mainly on the dynamics of a single disease. Recently,
however, the case of cooperative diseases, where positive feedback between
multiple infections can lead to more violent progressions, has gained
increasing attention (see publications cited in \cite{CGCG13}). One of the most
fundamental and interesting questions is whether cooperation in co-infections
can change the outbreak from being a continuous (\textquotedblleft second
order\textquotedblright) to a discontinuous (\textquotedblleft first
order\textquotedblright) phase transition because the amount of time one has to
enact countermeasures depends critically on the transition type. As the
favorability of conditions passes the transition point, the epidemic grows
continuously in a second order transition, whereas it can take over a
population explosively in a first order transition.

Recently, Chen, Ghanbarnejad, Cai, and Grassberger (CGCG)~\cite{CGCG13},
introduced a mean-field model of cooperative co-infection in a spatial
homogeneous (\textquotedblleft well stirred\textquotedblright) situation by
generalizing the \textquotedblleft
susceptible-infected-removed\textquotedblright (SIR) model~\cite{kerMcK1927}.
Assuming the symmetry between two diseases, CGCG derived a set of ordinary
differential equations describing the reactions of their model which then was
integrated numerically to find out the properties of the final steady states
where all disease activity has died out. As one of their main results, CGCG
report that their numerical results indicate the presence of a first-order
phase transition. Note though, that in non-equilibrium systems, the notion of a
first-order transition is not unique, and different definitions are being used
in the literature. The most stringent definition that is closest in spirit to
an equilibrium first-order transition in the sense of the traditional Ehrenfest
classification is one that requires the coexistence between phases at a
nonequilibrium transition too. However, the term first-order transition is also
often being used for discontinuous non-equilibrium transitions without
phase-coexistence, e.g. for a discontinuous transition at a spinodal point.
Perhaps somewhat judiciously, we adopt here the stringent definition and refer
to a discontinuous transition only as a first-order transition when there is
phase-coexistence and we refer to a discontinuous transition at a spinodal
point as a spinodal transition.

 In a non-equilibrium steady state one does not have a "tool" such as a free energy
potential to properly capture a first-order transition with a spatially homogeneous model,
and hence one has to resort to other approaches. For example, one can discuss the first-order
transition in terms of the drift velocity of a moving interface between both
phases (provided that this interface is well defined in the sense that it is
not a too rugged fractal), as we will do in the present paper. This velocity
has to be zero at the transition point \cite{HHL08}.

Here, we will argue that CGCG's model coincides in mean-field theory with the
spatially homogeneous limit of the extended general epidemic process (EGEP)
that we, Janssen, M\"{u}ller and Stenull (JMS) \cite{JMS04}, introduced some 10
years ago to discuss tricritical and first-order behavior in percolation
processes. Recently, the EGEP has been studied numerically by Bizhani, Paczusky
and Grassberger~\cite{BPG12}. This numerical study has verified the existence
of first-order transitions in spatial dimensions larger than two as predicted
by JMS.

Given the recent results by CGCG, we think that it is in order to revisit the
EGEP. After briefly reviewing the EGEP, we study the homogenous mean-field
theory that applies to both the CGCG model and the EGEP. The set of
differential equations defining the EGEP can be reduced to an equation of
motion for the removed (dead or immune) individuals. The temporal behavior of
the process and the properties of the ultimate final steady states can then be
understood simply by discussing the analytical behavior of the production-rate
for the fraction of the removed constituents, without the need for numerical
integrations. It turns out hat the discontinuous transition found by CGCG is a
spinodal transition.

\section{The EGEP}
We briefly review the EGEP  to provide background information and to establish
notation. The standard GEP \cite{Mol77,Bai75,Mur89,Gra83}, assumed to take
place on a \textit{d}-dimensional lattice, is described with help of the
reaction scheme
\begin{subequations}
\label{GEP}%
\begin{align}
S(\mathbf{r})+X(\mathbf{r^{\prime}})\quad &
\overset{\kappa}{\longrightarrow}\quad
X(\mathbf{r})+X(\mathbf{r^{\prime}})\,,\label{reactA}\\
X(\mathbf{r})\quad &  \overset{\lambda}{\longrightarrow}\quad Z(\mathbf{r})\,,
\label{reactB}%
\end{align}
with reaction rates $\kappa$ and $\lambda$. $S$, $X$, and $Z$ respectively
denote susceptible, infected, and removed individuals on nearest neighbor sites
$\mathbf{r}$ and $\mathbf{r^{\prime}}$ of a $d$-dimensional lattice. A
susceptible individual may be infected by an infected neighbor with rate
$\kappa$ [reaction~(\ref{reactA})]. By this mechanism the disease (henceforth
also called the agent) spreads diffusively. Infected individuals are removed
with a rate $\lambda$ [reaction~(\ref{reactB})]. JMS extended the
GEP-reaction-scheme by introducing \emph{weak }(or \emph{touched}) individuals
$Y$. Instead of being infected right away by an agent, any susceptible
individual may be weakened with a reaction rate $\mu$ by such an encounter.
When the disease passes by anew, a weakened individual is more susceptible to
attracting the disease and gets sick with a rate $\nu>\kappa$. Therefore, the
EGEP is described by the additional reactions
\end{subequations}
\begin{subequations}
\label{EGEP}%
\begin{align}
S(\mathbf{r})+X(\mathbf{r^{\prime}})\quad &
\overset{\mu}{\longrightarrow}\quad
Y(\mathbf{r})+X(\mathbf{r^{\prime}})\,,\label{react C}\\
Y(\mathbf{r})+X(\mathbf{r^{\prime}})\quad &
\overset{\nu}{\longrightarrow}\quad
X(\mathbf{r})+X(\mathbf{r^{\prime}})\,. \label{react D}%
\end{align}
Note, that these two reactions introduce a kind of cooperation through repeated
encounters of agents which then can lead to instabilities.

Denoting by $s,$ $w,$ $n,$ $m$ the fractions of the total population of the
constituents $S,$ $Y,$ $X,$ and $Z$, respectively, the reaction-equations read
\end{subequations}
\begin{subequations}
\label{meanfield}%
\begin{align}
\dot{s}(t,\mathbf{r})  &  =-(\kappa+\mu)s(t,\mathbf{r})\bar{n}(t,\mathbf{r}%
)\,,\label{meanS}\\
\dot{w}(t,\mathbf{r})  &  =\bigl(\mu s(t,\mathbf{r})-\nu w(t,\mathbf{r}%
)\bigr)\bar{n}(t,\mathbf{r})\,,\label{meanY}\\
\dot{n}(t,\mathbf{r})  &  =\bigl(\kappa s(t,\mathbf{r})+\nu w(t,\mathbf{r}%
)\bigr)\bar{n}(t,\mathbf{r})-\lambda n(t,\mathbf{r})\,,\label{meanX}\\
\dot{m}(t,\mathbf{r})  &  =\lambda n(t,\mathbf{r})\,, \label{meanZ}%
\end{align}
where%
\end{subequations}
\begin{equation}
\bar{n}(t,\mathbf{r})=\frac{1}{2d}\sum_{\mathbf{r^{\prime}}}^{nn(\mathbf{r}%
)}n(t,\mathbf{r^{\prime}})=\frac{1}{\lambda}\frac{\partial}{\partial t}\bar
{m}(t,\mathbf{r})\,.
\end{equation}
Here, $\sum_{\mathbf{r^{\prime}}}^{nn(\mathbf{r})}\cdots$ denotes summation
over the $2d$ nearest neighbors of site $\mathbf{r}$. Note, that in the
homogeneous limit, the first three reaction-equations of the scheme
(\ref{meanfield}) are identical to the reaction-equations of CGCG. At each
lattice site there is the additional constraint
\begin{equation}
s+w+n+m=1\,. \label{constraint}%
\end{equation}

Equations~(\ref{meanS}) and (\ref{meanY}) are readily integrated. Using the
initial conditions $s(0,\mathbf{r})=s_{0}(\mathbf{r})$, $w(0,\mathbf{r}%
)=w_{0}(\mathbf{r})$, $m(0,\mathbf{r})=0$, we obtain
\begin{equation}
s(t,\mathbf{r})=s_{0}(\mathbf{r})\exp\bigl(-\rho\bar{m}(t,\mathbf{r})\bigr)
\end{equation}
and
\begin{align}
w(\mathbf{r},t)  &  =w_{0}(\mathbf{r})\exp\bigl(-\nu\bar{m}(t,\mathbf{r}%
)\bigr)\nonumber\\
&  +\frac{\mu}{\nu-\rho}s_{0}(\mathbf{r})\Big\{\exp\bigl(-\rho\bar
{m}(t,\mathbf{r})\bigr)
\nonumber\\
&-\exp\bigl(-\nu\bar{m}(t,\mathbf{r})\bigr)\Big\}\,,
\label{simplesol}%
\end{align}
where we have defined $\rho=\kappa+\mu$. The time scale $\lambda$ has been set
to unity for simplicity. Equation~(\ref{meanZ}) together with the
constraint~(\ref{constraint}) finally leads to the mean field equation of
motion for the removed individuals of the EGEP:
\begin{align}
\dot{m}(t,\mathbf{r})  &  =1-m(t,\mathbf{r})-w_{0}(\mathbf{r})\exp
\bigl(-\nu\bar{m}(t,\mathbf{r})\bigr)\nonumber\\
&  -s_{0}(\mathbf{r})\Big\{\frac{\nu-\kappa}{\nu-\rho}\exp\bigl(-\rho\bar
{m}(t,\mathbf{r)}\bigr)
\nonumber\\
&-\frac{\rho-\kappa}{\nu-\rho}\,\exp\bigl(-\nu\bar
{m}(t,\mathbf{r)}\bigr)\Big\}\,\,. \label{GGEPstate}%
\end{align}

\section{Homogeneity and spinodal transition}
First we consider the spatially homogeneous case. We use the initial values
$s_{0}=1-\varepsilon$, $w_{0}=r\varepsilon$, implying $n_{0}=(1-r)\varepsilon$.
To establish contact with the notation of CGCG, we set $\rho=2\alpha$,
$\kappa=\alpha\delta$, $\nu=\alpha c$, where $c$ is a measure of cooperativity.
For convenience, we include the factor $\alpha$ in the definition of the
fraction $m$ of removed individuals, and set
\begin{align}
\alpha m \rightarrow m\,.
\end{align}
We make the simplifying replacements
$\varepsilon/(1-\varepsilon)\rightarrow\varepsilon$ and $\alpha(1-\varepsilon
)\rightarrow\alpha$ as we can because we are interested in small $\varepsilon$,
$\varepsilon\sim10^{-3}$ or so. Using the CGCG-value $\delta=1$ (in addition
CGCG use $r=1/2$), the equation of motion reads
\begin{subequations}
\label{eqofmot}%
\begin{align}
\frac{\dot{m}}{\alpha}  &  =R(m)=G(m,c)-\frac{m}{\alpha}+\varepsilon
\bigl(1-r\mathrm{e}^{-cm}\bigr)\,,\label{diffeq}\\
G(m,c)  &  =\Big\{1-\frac{c-1}{c-2}\mathrm{e}^{-2m}+\frac{1}{c-2}
\mathrm{e}^{-cm}\Big\}\,. \label{rate}%
\end{align}
\end{subequations}
This differential equation can be solved by a simple integration. Using $m=0$
for $t=0$, we obtain the inverse function
\begin{equation}
t=I(m):=\int_{0}^{m}\frac{dm^{\prime}}{\alpha R(m^{\prime})}
\end{equation}
of $m(t)$. From this result, one can readily extract analytical prediction for
the time-dependent quantities in the numerical study of CGCG.

\begin{figure}[ptb]
\centerline{\includegraphics[width=4.4cm]{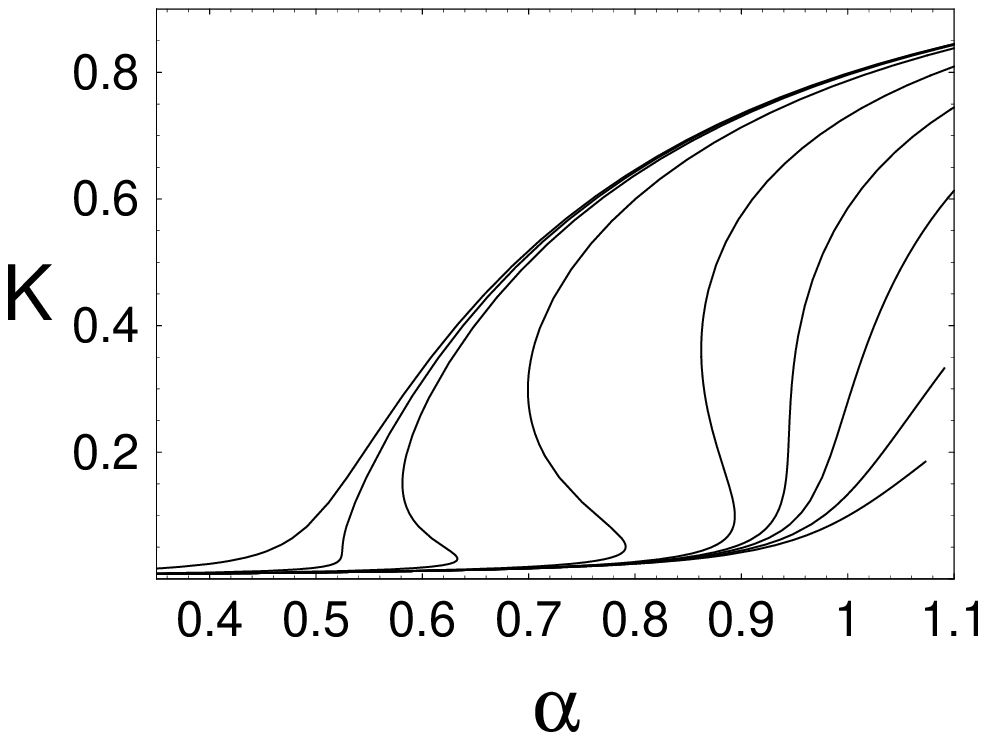}
\includegraphics[width=4.4cm]{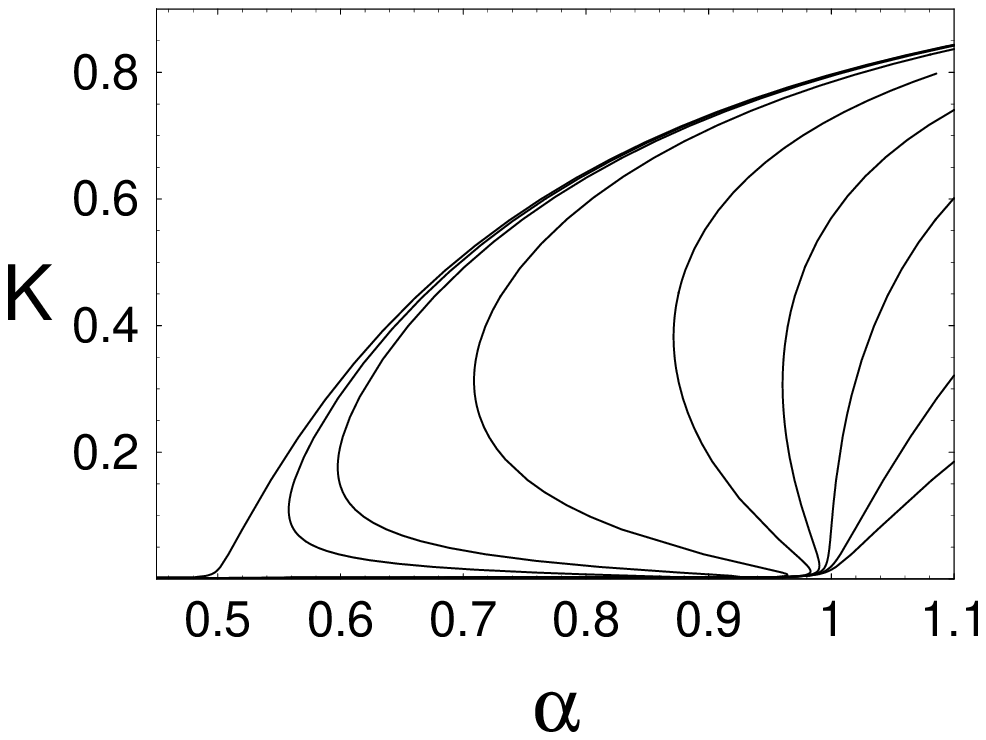}}
\caption{Order
parameter $K$ plotted against $\alpha$ for $\varepsilon=0.005$ (top left) and
$\varepsilon=10^{-4}$ (top right). The collection of curves in each diagram
corresponds to $c=\infty,160,60,15,5,3,2,1,0.1$ (same values as used in the
simulation of CGCG) from left to right.
}
\label{CollOrdP}
\end{figure}

Here, however, we are mainly interested in the final steady state
with $n(t\rightarrow\infty)=0$, where all activity has died out.
This state is determined by the $m=m_{\infty}:=m(t\rightarrow\infty)$ that solves the
equation $R(m_{\infty})=0$. To simplify the discussion of our
analytical results and their comparison to the numerics
of CGCG, we switch from $m$ as our order parameter to the fraction
of converted susceptibles
\begin{equation}
K=1-s=1-s_{0}\exp\bigl(-2m\bigr)
\end{equation}
as used by CGCG. Note that this mapping from $m$ to $K$ is one-to-one.
Figure~\ref{CollOrdP} displays a collection of curves of
$K=K_{\infty}:=K(t\rightarrow\infty)$ as functions of
$\alpha$ with parameter $c$ for $r=1/2$ and small $\varepsilon$,
$\varepsilon=5\cdot10^{-3}$ and $\varepsilon=10^{-4}$. Comparing
Fig.~\ref{CollOrdP} with Fig.\thinspace2 of Ref.~\cite{CGCG13}, we
note that the numerical curves of CGCG are implied in our analytical
curves.

\begin{figure}[ptb]
\centerline{\includegraphics[width=5.5cm]{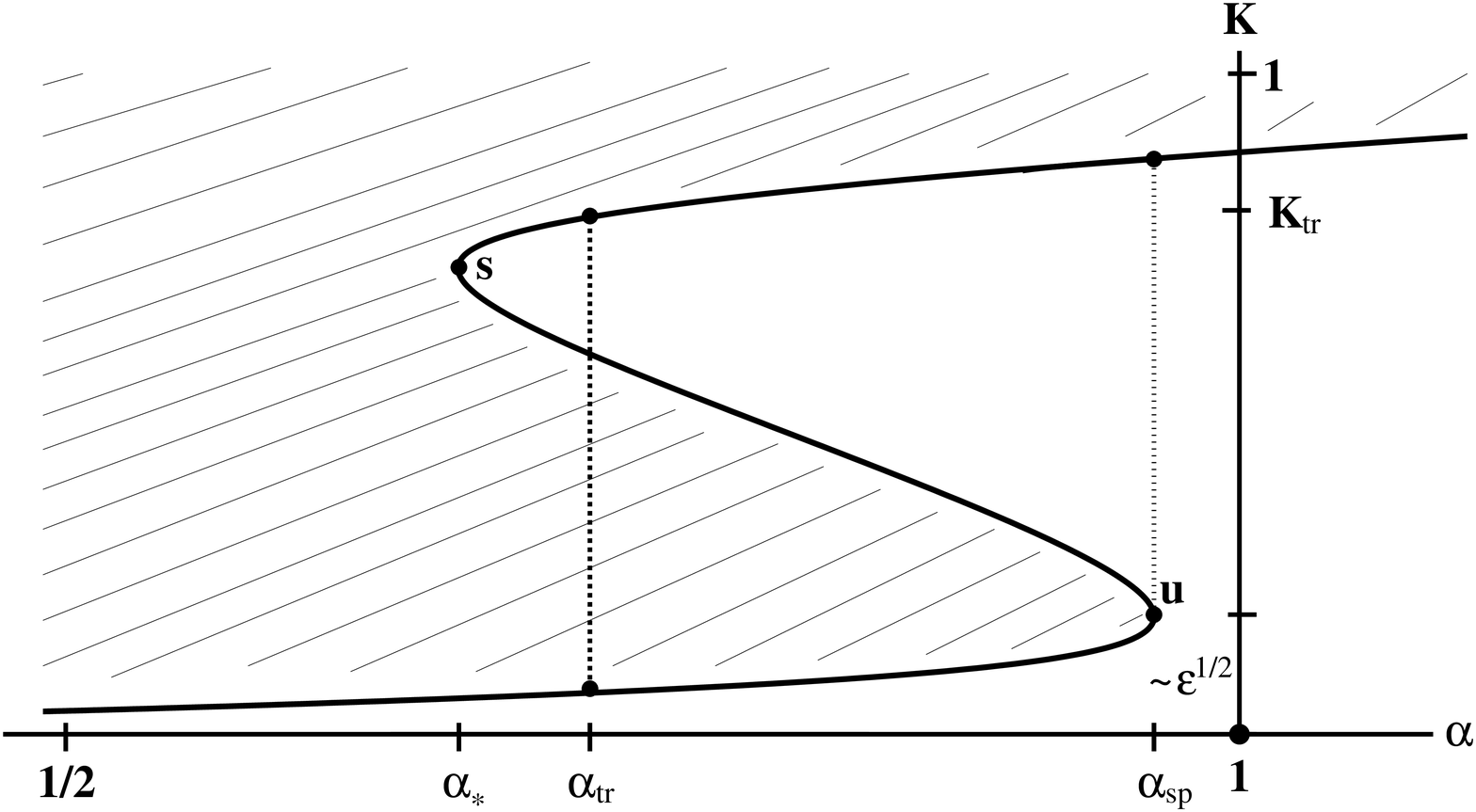}} \caption{Schematic sketch
of the order-parameter as a function of $\alpha$.}
\label{sketch}%
\end{figure}

We think that it is worthwhile to discuss the order parameter curves in some
more detail. Figure~\ref{sketch} shows a schematic sketch of a generic order
parameter curve for some given $c$. Only those parts of the curve that have a
positive gradient are locally stable. The part between the spinodal points $s$
and $u$ is an unstable branch. In the hatched part of the figure, the rate
$R(m)$ is negative and therefore nonphysical in the homogeneous case. Let's
suppose we want to follow a process with a given set of parameters from its
beginning to its end. Every such process starts with $m=0$, i.e.,
$K=\varepsilon$, and its order parameter is the final $K=K_{\infty}$.
Each process with fixed $c$ and $\alpha$ corresponds to an upward
directed vertical line in Fig.~\ref{sketch}. If $\alpha < \alpha_{sp}$, the
upper part of the curve is inaccessible to the process and its final $K$ lies
on the lower branch. If $\alpha > \alpha_{sp}$, the final $K$ lies on the upper
branch. Thus, the order parameter jumps right at the spinodal point $u$, where
the parameter $\alpha$ takes the value $\alpha_{sp}$. This jump is in
perfect agreement with the discontinuities produced by the numeric integration
procedure of CGCG. Note that the numerical curves shown Fig.\thinspace2 of Ref.~\cite{CGCG13} and our analytical curves can literally be superimposed, including the jumps. CGCG interpret these discontinuities as first-order phase transitions with thresholds $\alpha=\alpha(c)$. We learn
here, though, that they are transitions at a spinodal point, the point where
the lower locally stable part of the curve becomes unstable. Since the
reactions are completely irreversible, the upper spinodal point $s$ with
corresponding $\alpha=\alpha_{\star}$ is irrelevant, and no hysteresis appears.
$\alpha_{tr}$ pertains to the EGEP when spatial inhomogeneity is permitted, see
further below, where we will argue that true first-order transition appears in
the EGEP at values $\alpha=\alpha_{tr}(c)$ in the interval between $1/2 <
\alpha < 1$ even in the limit $\varepsilon \to 0$ where all the
$\alpha_{sp}(c)$ tend to $1$.

\begin{figure}[ptb]
\centerline{\includegraphics[width=5.5cm]{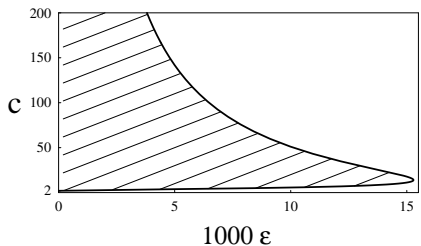}} \caption{The range of
parameters (shaded area) for which there is a discontinuous transitions.}
\label{Epsoverc}
\end{figure}

In addition to the order-parameter curves, all the other numerical results by
CGCG absolutely agree with and in fact can be extracted from the analytical
results presented in Ref.~\cite{JMS04} including the mean-field behavior near
the tricritical point $c=2$, $\alpha=1$ and the critical line $c<2$,
$\alpha=1$. We note that the limit $c\rightarrow\infty$ leads back to the
ordinary GEP with a continuous transition now at $\alpha=1/2$.

Beyond these established results, we think it is interesting to calculate the
range of values of the cooperativity $c$ over which the discontinuous behavior
seen in the numerics of CGCG exists. The end points of this range are defined
by the inflection point with vertical tangent of the order-parameter curves.
Hence, to determine these points,  we seek the solutions of
$R(m)=dR(m)/dm=d^{2}R(m)/dm^{2}=0$, and we obtain
\begin{subequations}
\label{range}%
\begin{align}
\varepsilon &  =\Big[\frac{c}{2(c-2)}\ln\Big(\frac{c^{2}}{4(c-1)}%
\Big)+\frac{c+2}{4}\Big]\Big[\frac{4(c-1)}{c^{2}}\Big]^{c/(c-2)}
\label{range 1}
\nonumber\\
&-1\,,
\\%
\alpha &  =\frac{2}{c}\Big[\frac{c^{2}}{4(c-1)}\Big]^{c/(c-2)}, \quad
K=1-\Big[\frac{4(c-1)}{c^{2}}\Big]^{2/(c-2)}\,.\label{range 2}
\end{align}
Figure~\ref{Epsoverc} shows a plot of solution~(\ref{range 1}). We observe that
discontinuous transitions occur only if $\varepsilon$ is smaller than a maximum
value, $\varepsilon\leq\varepsilon_{M}\approx0.016$. In the limit
$\varepsilon\rightarrow0$ the range of discontinuous transitions extends from
$c=2$ to $c=\infty$.

\section{Inhomogeneity and first order transition}

Next, we consider the spatially inhomogeneous case to elucidate the spreading
behavior and, in particular, the conditions for phase coexistence. When going
beyond the spatially homogenous limit, the CGCG model is more complex than the
EGEP in that corresponds to two distinct GEPs cooperating with one another
whereas the EGEP corresponds to one GEP cooperating with itself. We will in the
remainder set aside the CGCG model focus here on the EGEP where a description
in terms of a single "field", i.e., the fraction of converted individuals
$m(t,\mathbf{r})$, is guaranteed to work.

Assuming at the lattice constant $a$ is small compared to the length of spatial
variations of $m(t,\mathbf{r})$, we approximate
\end{subequations}
\begin{equation}
\bar{m}(t,\mathbf{r})=\frac{1}{2d}\sum_{\mathbf{r^{\prime}}}^{nn(\mathbf{r}%
)}m(t,\mathbf{r^{\prime}})\approx m(t,\mathbf{r})+\frac{a^{2}}{2d}\nabla^{2}%
m(t,\mathbf{r})\,.
\end{equation}
With this approximation, a gradient expansion of the equation of motion, Eq.\
(\ref{GGEPstate}), results in the reaction-diffusion equation
\begin{equation}
\frac{\dot{m}}{\alpha}=R(m)+\lambda G^{\prime}(m)\nabla^{2}m\,,
\label{ReactDiffus}%
\end{equation}
where $\lambda$ is some combination of constants and the prime $G^{\prime }$
denotes the derivative of $G$ with respect to $m$. Note that according to
Eq.~(\ref{GGEPstate}),  $R$ and  $G$  depend on the initial distribution of the
susceptibles and agents, and therefore are, in general,functions of the spatial
position. Because we are interested in the final steady state determined by
$\dot{m}=0$, we focus on the steady state condition
\begin{equation}
\lambda\nabla^{2}m+V^{\prime}(m)=0\,, \label{DiffBdg}%
\end{equation}
where $V(m;\alpha,c)$ is a \textquotedblleft potential\textquotedblright\
defined by the integral
\begin{equation}
V(m;\alpha,c)=\int_{0}^{m}dm^{\prime}\frac{R(m^{\prime};\alpha,c)}{G^{\prime}(m^{\prime};c)}\,,
\label{Potential}%
\end{equation}
where we have explicitly displayed the dependence on the parameters $\alpha$
and $c$. Next, we determine the point $\alpha_{tr}$ where the disease changes
from endemic to pandemic behavior, that is where the spreading, initiated by an
completely infected planar slab, begins to evolve but stops in a formerly
non-infected region ($\varepsilon=0$). To this end, we seek a solution of the
steady state in the form of an interface between the two phases with
order-parameters $m=m_{1}$ and $m=m_{2}$ determined by $R(m_{1})=R(m_{2})=0$
that is planar and perpendicular to the coordinate $x$. Then the steady state
condition (\ref{DiffBdg}) leads to
\begin{equation}
\frac{d}{dx}\Big\{\frac{\lambda}{2}\Big(\frac{dm}{dx}\Big)^{2}+V(m)\Big\}=0\,.
\label{EnCons}%
\end{equation}
which is analogous in form to a typical equation of motion with
\textquotedblleft energy\textquotedblright conservation in classical mechanics.
Because $dm/dx\rightarrow0$ deep in each phase, we have $V(m_{1})=V(m_{2})$,
which leads to
\begin{equation}
V(m_{2};\alpha,c)-V(m_{1};\alpha,c)=\int_{m_{1}}^{m_{2}}dm\frac{R(m,\alpha,c)}{G^{\prime}(m,c)}=0\,,
\label{EquilibrCond}%
\end{equation}
as the condition of coexistence between the two phases. Note that this
coexistence condition (\ref{EquilibrCond}) is completely analogous to the equal
area construction well known from the van-der-Waals equation in thermostatics.
Together with $R(m_{1},\alpha,c)=R(m_{2},\alpha,c)=0$, this condition uniquely
determines the locus $\alpha=\alpha_{tr}(c)$ and discontinuity $m_{tr}(c)$ of a
true first-order transition. Table~\ref{tab:trans} compiles values of
$\alpha_{tr}(c)$ and $m_{tr}(c)$ for $\varepsilon=0$. Note that for
$\varepsilon=0$, the spinodal point is for all $c$ at $\alpha_{sp}=1$.
\begin{table}
\begin{tabular}{c|ccccc}
$c$ & $3$ & $5$ & $15$ & $60$ & $160$\\
\hline
$\alpha_{tr}$ & $0.964$ & $0.882$ & $0.724$ & $0.609$ & $0.566$\\
$m_{tr}$ & $0.251$ & $0.337$ & $0.274$ & $0.152$ & $0.095$
\end{tabular}
\caption{Values of the transition point $\alpha_{tr}(c)$ and the discontinuity
$m_{tr}(c)$ for $\varepsilon=0$.} \label{tab:trans}
\end{table}
The usual integration of the "energy-conservation law", Eq.~(\ref{EnCons}),
finally leads to the density profile between the two phases.

To further evaluate the coexistence condition analytically, we expand $R(m)$
for $\varepsilon=0$ in powers of $m$,
\begin{equation}
R(m)=-\tau m+\frac{\sigma}{2}m^{2}-\frac{g}{6}m^{3}  \,,
\label{rateSch}%
\end{equation}
where we have discarded higher order terms, where $G^{\prime}(m)$ was replaced
by a constant included in $\lambda$, and where $\tau=1/\alpha-1$, $\sigma=c-2$,
and $g=c^{2}+2c-4$.  With these approximations, the potential defined in
Eq.~(\ref{Potential}) becomes
\begin{equation}
V(m)=-\frac{\tau}{2}m^{2}+\frac{\sigma}{6}m^{3}-\frac{g}{24}m^{4} \, .
\label{PotNaeh}%
\end{equation}
Note that our approximation, although certainly not quantitatively correct,
comprises all qualitative aspects of the EGEP.

Now, we return to the coexistence condition and its evaluation.
Equation~(\ref{EquilibrCond}) in conjunction with the fact that $m_{1}$ and
$m_{2}$ are solutions of $V^{\prime}(m) \sim R(m)=0$  leads to
\begin{align}
\frac{\tau_{tr}}{g}  &  =\frac{1}{3}\Big(\frac{\sigma}{g}\Big)^{2}\,, \qquad
m_{1}    = 0\,,\qquad m_{2}    = 2\frac{\sigma}{g}\,,
\end{align}
for the transition point and the order-parameters of the two phases. These
results imply that the discontinuity $d_{tr}=m_{2}-m_{1}$ at the first order
transition is
\begin{equation}
d_{tr}=2\frac{\sigma}{g}\,.
\label{FirstOrder-Disc}%
\end{equation}

From our approximation, Eq.~(\ref{rateSch}), we can also extract information
about the spinodal transition. Our discussion above (see Fig.~\ref{sketch})
implies that the rate $R(m)$ (\ref{rateSch}) fulfils the conditions
$R(m_3)=R^{\prime}(m_3)=R(m_4)=0$ at the spinodal transition point $\tau
=\tau_{sp}=0$, where $m_{3}$ and $m_{4}$ denote the order-parameter values at
that point. It follows that
\begin{equation}
m_{3}  = 0\,,\qquad m_{4}=3\frac{\sigma}{g}\label{SpinodTrans}
\end{equation}
The spinodal discontinuity $d_{sp}=m_{4}-m_{3}$ therefore is
\begin{equation}
d_{sp}=3\frac{\sigma}{g} \,. \label{SpinodDisc}
\end{equation}
in agreement with our discussion of the spatially homogenous case.

Having discussed steady state and coexistence properties, we would
like to close by mentioning that approximate analytical results for the
time-dependence of the order-parameter are available. In Ref.~\cite{JMS04}, we
have studied if and under which conditions a percolating interface can emerge
and spreads from a plane-like source, i.e., an entire infected
slab in the region $-x/a \gg 1$. Indeed, the reaction-diffusion equation
(\ref{ReactDiffus}) has such a traveling-front-solution. With the notation used
here, this solution reads
\begin{equation}
m(x,t)=A\Big\{1-\tanh[b(x-vt)]\Big\}\,,\label{TravWav}%
\end{equation}
where
\begin{subequations}
\label{TrW-Par}%
\begin{align}
A &
=\frac{3}{4}\bigg(\sqrt{\Big(\frac{\sigma}{g}\Big)^{2}-\frac{8\tau}{3g}}+\frac{\sigma}{g}\bigg)\,,\qquad
b=\sqrt{\frac{g}{12\lambda}}A\,,\label{TrW-Par1}\\
v &
=\frac{3\alpha\sqrt{3g\lambda}}{4}\bigg(\sqrt{\Big(\frac{\sigma}{g}\Big)^{2}-\frac{8\tau}{3g}}
-\frac{\sigma}{3g}\bigg)
\,.\label{TrW-Par2}%
\end{align}
\end{subequations}
The positivity of $\dot{m}$ requires that $v\geq0$. At the first-order
transition, $v$ vanishes, whereas at the spinodal point $\tau=0$ it is
\begin{equation}
v_{sp}=3\alpha \sqrt{\lambda/2}\;\frac{\sigma}{g}>0
\end{equation}
which shows that there is no phase-coexistence at this point. The
percolating front of the disease propagates with a finite velocity
(explosively) under spinodal conditions. Near the first-order transition, $\tau
\leq \tau_{tr}$, the velocity behaves as $v\sim(\tau_{tr}-\tau)$, which implies
a depinning transition with a mean-field exponent $\beta =1$.

\section{Concluding remarks}
The present letter serves three purposes. First, it comments on the recent
letter by CGCG. It demonstrates that the jumps in the numerical curves by CGCG
have to be interpreted basically as a spinodal rather than a true first-order
phase transition with phase coexistence. Second, it derives the conditions for
having a true first-order transition within the underlying model system and it
provides various results for physical observables at this transition such as
the magnitude of the order-parameter discontinuity etc. Third, it contributes
to the basic understanding of discontinuous percolation transitions, an area of
statistical physics that has recently enjoyed expanding activity
\cite{AGKSZ14}.

To avoid any potentially lingering confusion, we would
like to stress here that first-order and spinodal transitions in
non-equilibrium systems are both discontinuous phase transitions.
Both display a discontinuity of the same order parameter at the
transition, however, both start with different initial states. The
following important difference between them arises if one admits
inhomogeneous states:  At a first-order transition, one has $2$
distinct phases coexisting in the final steady state with a
stationary interface between them. At a spinodal transition, a
spatially homogeneous initial state undergoes a discontinuous
transition to a new homogeneous final state, and a initial
point-like seed of the contagion would spread out infinitely far
with finite velocity.

It is worth to emphasize some more the interesting connection between
disease spreading and a depinning-transition that emerged above.
When assuming a $(d-1)$-dimensional planar slab of infected
individuals as the initial state of the EGEP, the mean-field picture
that we saw was the following: Below the threshold $\alpha_{tr}$,
the front of the epidemic process progresses diffusively until it
stops before a macroscopically large area has become infected. Right
at $\alpha_{tr}$, the front progresses diffusively. When the process
stops, a macroscopically large area has become infected with an
interface between the infected and non-infected regions that is
shaped, on average, as described above. This final state is what we
have been alluding to as phase-coexistence. Above $\alpha_{tr}$, the
front progresses ballistically, and the infection takes over the
entire system. The qualitative change in the dynamics of the
interface right at $\alpha_{tr}$ clearly has the character of a
critical depinning-transition.

In this letter, we have restricted our discussion to mean-field theory.
In a future publication \cite{JSt16}, we will derive a
field-theoretic interface model for the EGEP, and we will discuss the
first-order transition and the accompanying continuous depinning-transition of
the percolating front of the contagion beyond mean-field theory. Giving up the
mean-field condition, the CGCG-model and the EGEP behave differently. The
CGCG-model corresponds to two cooperatively coupled ordinary GEP's
\cite{CCGG15} which leads to a hybrid transition with both first-order and
universal second-order phase-transition aspects and a very rich phenomenology.
The EGEP, on the other hand, corresponds to only one GEP cooperating with
itself, and it leads to a phase diagram with a line of true first-order
transitions with phase coexistence as the threshold of spreading that is
separated from a line of continuous ordinary percolation transitions by a
tricritical point. The critical depinning-transition accompanying the
discontinuous transition has a rough interface which seems to belong to the
universality class of critically pinned interfaces in isotropic random media
\cite{BPG12}.

Recently, there has been considerable interest in the
literature in hybrid phase transitions that show both first-order
and second-order behavior \cite{AGKSZ14}. Though we have not yet
studied any of these systems in detail, we note here the intriguing
possibility that such hybrid transitions, in particular in random
and interdependent networks without short loops (locally tree-like
and therefore with generic mean-field behavior) \cite{GoDoMe06, DoGoMe08,
BuPaPaStHa10, PaBuHa10, PaBuHa11, GaBuStHa11}, are typically
associated with spinodal transitions \cite{SoGrPa11, SoBiChGrPa12,
Gra15}.

\acknowledgments  This work was supported by the NSF under No.~DMR-1104701 (OS)
and No.~DMR-1120901 (OS). We thank Peter Grassberger for numerous valuable
remarks, discussions, and for showing us as yet unpublished work.

\end{document}